\documentclass[doublecol]{epl2} 
\usepackage{graphicx}
\newcommand{\beq}{\begin{equation}}
\newcommand{\eeq}{\end{equation}}

\title{Searching and fixating: scale-invariance vs. characteristic timescales in attentional processes}
\author{D. P. Shinde \inst{1} \thanks{E-mail: \email{shinde@bose.res.in}} \and Anita Mehta \inst{1} \thanks{E-mail: \email{anita@bose.res.in}} \and R. K. Mishra \inst{2} \thanks{E-mail: \email{rkmishra@cbcs.ac.in}}}
\institute{                    
  \inst{1} Theory Department, S N Bose National Centre, Block JD Sector III, Salt Lake, Calcutta 700098, India\\
  \inst{2} Centre for Behavioural and Cognitive Sciences, University of Allahabad, India
}

\pacs{89.75.-k}{Complex systems}
\pacs{05.40.-a}{Fluctuation phenomena, random processes, noise, and Brownian motion}
\pacs{05.45.Tp}{Time series analysis}
\pacs{87.19.lt}{Sensory systems}

\abstract{In an experiment involving semantic search, the visual movements of sample populations subjected to visual and aural input were tracked in a taskless paradigm. The probability distributions of saccades and fixations were obtained and analyzed. Scale-invariance was observed in the saccadic distributions, while the fixation distributions revealed the presence of a characteristic (attentional) time scale for literate subjects. A detailed analysis of our results suggests that saccadic eye motions are an example of  Levy, rather than Brownian, dynamics.}

\begin{document}
\maketitle

While there has been extensive quantitative analysis of complexity in the natural world \cite{Mehta, Kanamori}, the quantitative analysis of complex cognitive systems is still in its infancy. We know that our eyes respond to verbal influences by looking at an object of speech, or by imagining its visual analogues. Psychologists have carried out experiments on eye movements during reading \cite{Rayner1,Rayner2,Buswell} and scene perception \cite{Henderson1,Henderson2} and suggested that visual mechanisms can be broadly divided into \textit{saccades} and \textit{fixations}. During a saccade, the visual system is involved in search, and cognitive processing is minimal, while when it is \textit{fixated} on an object, search is temporarily at an end, and cognitive processing is maximal. Recently, there has been some work on saccadic processes in isolation (i.e. when fixations are not taken into account); however, their conclusions are often divergent \cite{Mark, stephen}. In fact, saccades and fixations correspond to changing states of attention in a dynamical system, so that a realistic quantitative approach would examine each one, as well as how each one affects the other; to the best of our knowledge, this has not been done prior to the present study. Additionally, subjects who are given an overt task to perform may respond differently from those who are tracking images through unconscious processing; the second process might yield insights that are more '\ natural '\, and it is for this reason that the design of the experiment analyzed below involves a taskless paradigm. Last but not least, we examine the effect of literacy on our observations, since literacy may have a direct influence on attentional mechanisms affecting eye movement \cite{Itti}.

The link between attention and overt eye movements has long been debated \cite{Posner}; it is now generally accepted that attention is not focused on any object during saccades, as the system searches for its next fixation. For our purposes, since we are interested in quantifying frequencies of saccades and fixations during a language-mediated visual search, we assume that fixations indicate zones of attentional stability, while saccades involve random, relatively inattentive, search. Another issue involves the selection of a comprehensive sample, with varying literacy levels; for this, one needs to construct a task which relies as little as possible on formal education. The visual world eye-tracking paradigm \cite{Cooper} was used in an experiment whereby overt visual attention to different objects was measured when participants looked at a display and processed an auditory sentence about it. Participants are under \textit{no} instructional control, and are \textit{fully} unaware that they are performing a set task. Saccades and fixations during this natural search are influenced by both the emerging visual and linguistic representations, and their interactions \cite{Gonzalez,Macknik}. We underline that no previous study  has measured eye movements in illiterate subjects, with a view to examining their attentional responses.

\begin{figure}
\centering
\includegraphics[height=6cm, width=7cm]{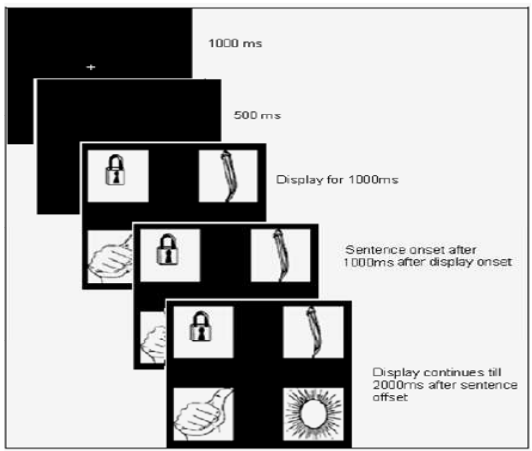}
\caption{Example of a trial sequence: the sample stimulus shows the phonological and semantic competitors of a target word. In this case the auditory sentence was : \textit{Usne aaj taara dekha}, (He saw a star today), where the phonological analogue to \textit{taara} (star) was \textit{taala} (lock), while the semantic analogue was the image of the sun. The other two images are distractors}
\label{fig.1}
\end{figure}

The display included four black and white line drawings of common objects. One of these objects was semantically related to the target spoken word in the auditory sentence while another was related phonologically, e.g. the words may have shared the first syllable. The other two objects were distractors. The participants included 36 illiterate and 32 literate subjects, selected from the Hindi-speaking population of Allahabad \cite{littest}. The spoken sentences were neutral, and the target word came after about 16 seconds on average from the sentence onset. Fig.~\ref{fig.1} illustrates a sample display. The fixation and saccadic eye movement data were collected and recorded with an SMI High Speed eye tracker running at a sampling rate of 1250 Hz; this recovered the X and Y coordinates of  a gaze with an accuracy of 0.01 degree, and the resulting data were binned using MATLAB. This was done for 35 trials per subject.

\begin{figure}
\centering
\includegraphics[width=.4\textwidth]{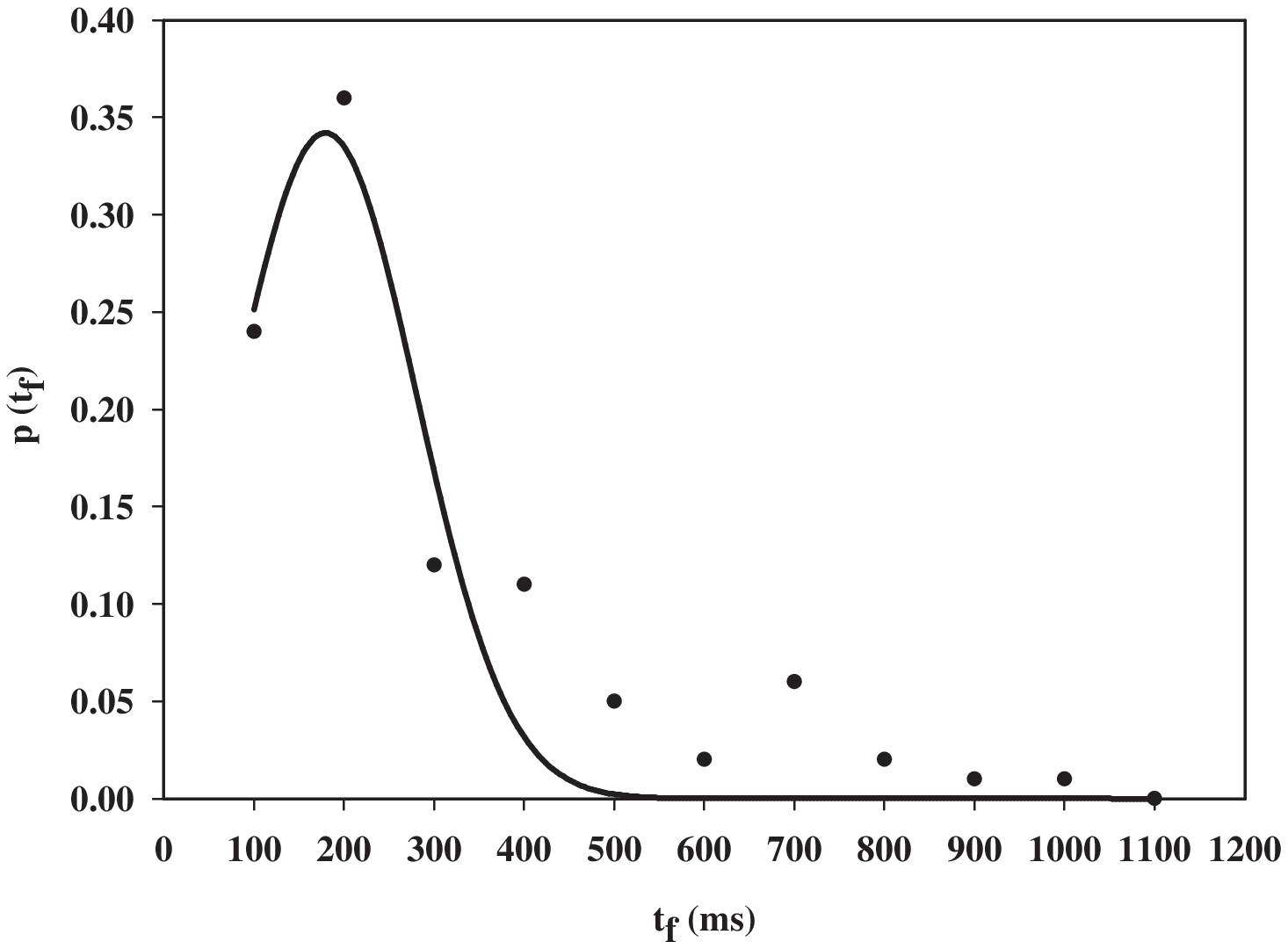}
\vskip 1mm
\includegraphics[width=.4\textwidth]{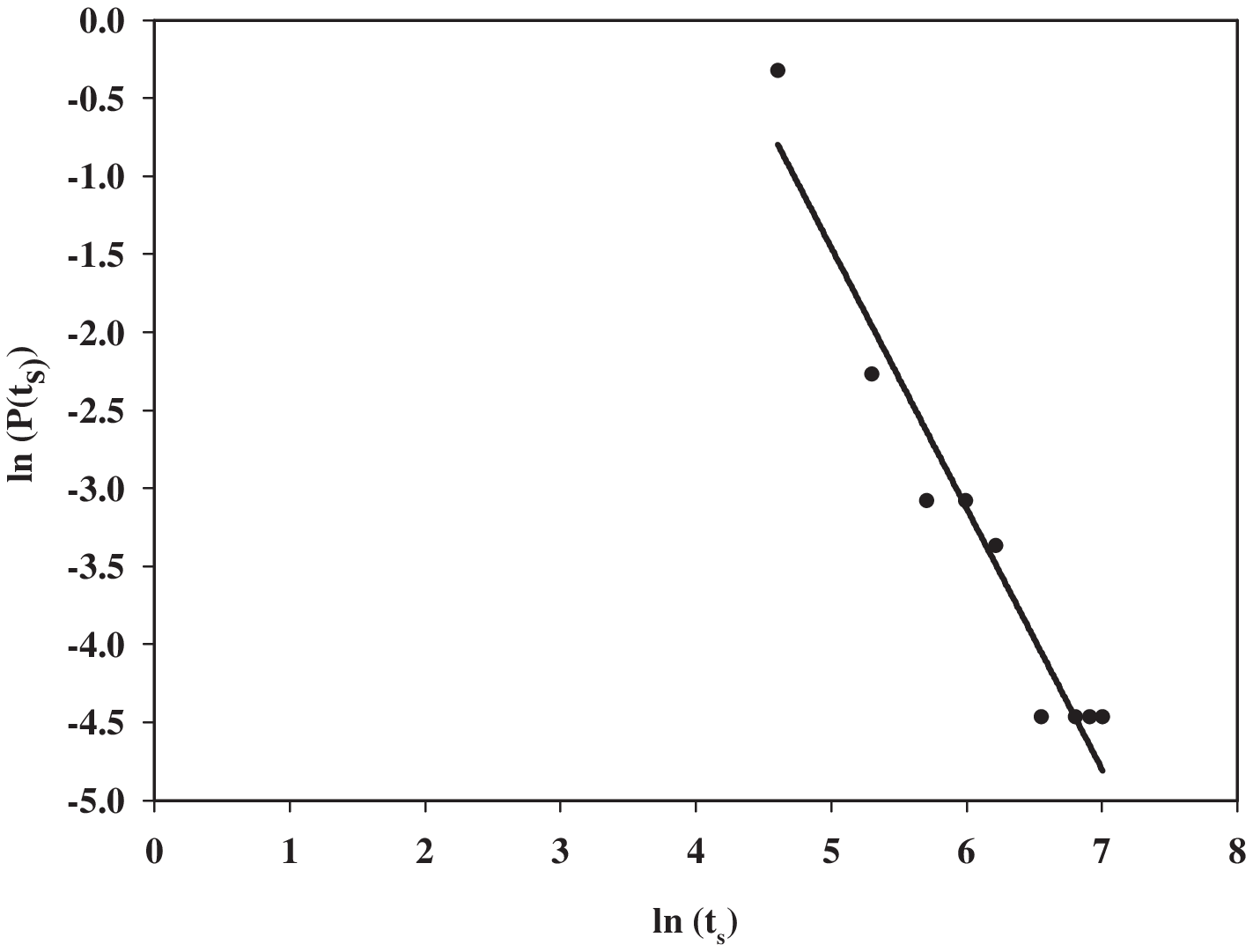}
\caption{Upper: The probability distribution $P (t _ f)$ of fixation time $t _ f$ for a typical literate subject. The peak of the distribution is at $179.12  \pm 16.14$ msec. Lower: The log of the probability distribution $P (t _ s)$ of saccadic time $t _s $ vs ln $(t _ s)$ for a typical literate, showing a power law with exponent $\beta=-1.67 \pm 0.14.$ }
\label{fig2}
\end{figure}

Saccades were identified following a velocity-based algorithm when the movement of the eye was greater than $30^{\circ}/sec$ in any direction from its current position. Fixations were identified with the stationarity of eye motion in any location,  for a minimum duration of 80 msec occurring between two saccades. For each of the subjects, the data were manually scanned, and probability distributions were computed to show the frequencies of fixation/saccade durations.
Overall, saccades appear to follow power-law distributions, $P(t _ s) = (t _ s)^{-\beta}$, where $t_{s}$ is saccadic time; averaging over all the subjects, we obtain the exponent $\beta=-1.77 \pm 0.23$, for both literate and illiterate subjects, with only about four of the illiterate subjects and one of the literate subjects showing significant deviations from such behavior. We observed however, that  illiterate subjects tended to spend relatively larger amounts of time on saccadic movements, even when power-law behavior was apparent overall. In Fig.~\ref {fig2}, we show the distributions of fixations and saccades for a typical literate subject.

\begin{figure}
\centering
\includegraphics[width=.4\textwidth]{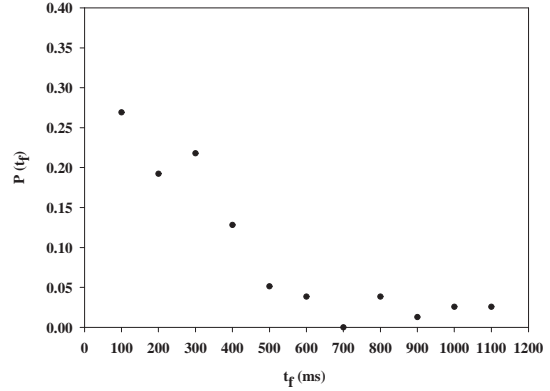}
\vskip 1mm
\includegraphics[width=.4\textwidth]{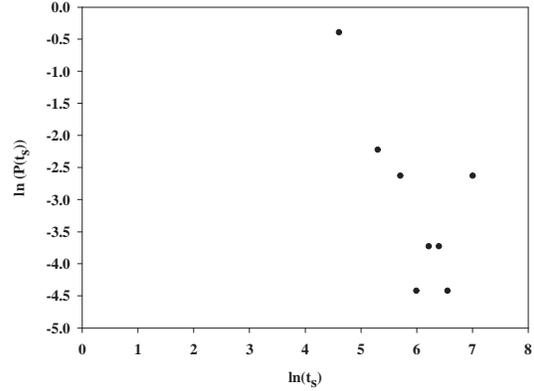}
\caption{Upper: The probability distribution $P (t _ f)$ of fixation time $t _ f$ for an extreme illiterate. No characteristic peak is discernible. Lower: The log of the probability distribution $P (t _ s)$ of saccadic time $t _s$ vs ln $(t _ s)$ for an extreme illiterate, where no power law is discernible. No lines have been fitted to the data for this reason.}
\label{fig3}
\end{figure}

The major difference between the two types appears in the distribution of fixations, which are marked by the presence of a characteristic time scale for the literate subjects (see Table ~\ref {table1}). A skew one-peaked distribution with a peak that occurs in the range 190-260 msec (see Table ~\ref {table1}), with long and occasionally complex tails, typifies the fixation distribution for literates; this suggests the presence of a typical attention-related peak when the subjects find meaningful correlations between a target image and the spoken word. For the illiterates, the distributions are less uniform. While some of them are rather similar to the skew 1-peaked distribution found in the literate case, most of them seem to exhibit power-laws, with no characteristic timescales, so that no fixation peaks are manifested (see Table ~\ref {table1}): significantly, in the latter case, there are high occurrences of large saccadic times. This seems to suggest that illiterate subjects show poorer cognitive processing, searching extensively and rarely fixating on their targets. The distributions for such an extreme case are shown in Fig.~\ref {fig3}. The position of the fixation (attentional) peak is absent, and no convincing line can be fitted to the data for the saccades either.

\begin{table}
\caption{Fixation time (in msec) distributions of 32 literates and 36 illiterates. Note that all 32 literates had fixation peaks, while only 5 out of 36 literates manifested fixation peaks.}
\label{table1}
\begin{center}
\begin{tabular}{ l | l | l | }
\hline
Literacy level&Serial No.&Peak value \\ \hline
{Literates}&6,9,21,22,25,29,30& $194.88 \pm 12.92$ \\ \cline{2-3}
&10,11,19,26,28,31&$221.36\pm7.12$ \\ \cline{2-3}
&1,3,4,7,8,14,24&$ 234.29\pm3.05$ \\ \cline{2-3}
&2,15,16,18,20,23,32&$243.84\pm2.62$ \\ \cline{2-3}
&5,12,13,17,27&$259.67\pm9.66 $ \\ \hline
{Illiterates}&5,4,10,14,23 & $247\pm23.57 $ \\ \cline{2-3}
&Others&did not fixate \\
\hline
\end{tabular}
\end{center}
\end{table}

In the remainder of this Letter, we focus on saccadic eye movement, which has attracted an immense amount of attention recently \cite{Macknik, Moshel1,Moshel2,Liang}.  There is an ongoing controversy on whether saccadic dynamics can be characterized as diffusive \cite{Mark} or hyperdiffusive \cite{stephen}. Traditional approaches to cognitive science had modeled language comprehension and visual search as ordinary diffusion \cite{Ratcliff, Palmer}; cognition was there viewed as a short-memory diffusion process that propagated as a linear function of time or stimulus-set size \cite{Horowitz,Wagenmakers}. More recent results have found that strong temporal correlations exist in cognitive behavior \cite{Orden}, suggesting that hyperdiffusion prevails. In their turn, hyperdiffusive processes may be classed as Gaussian or Levy, depending on whether the central limit theorem holds \cite{shlesinger1,shlesinger2}; this corresponds to whether or not it is temporally uncorrelated \cite{shlesingerwest, stanley}. Typically, a process $x(t)$ is characterised by two exponents: the diffusion exponent $\delta$,  obtained from $x(t) \sim  t^{\delta}$, and the Hurst exponent $H$, obtained from $V(t) =  <|x(t) - x(0)|^{2}> \sim  t^{2 H}$, where $V(t)$ is the mean-squared displacement of the process $x(t)$. When $\delta = H$ , the process is Gaussian; else, it is  Levy \cite{Zanette,shlesinger1,shlesinger2}. 

\begin{figure}
\centering
\includegraphics[width=.4\textwidth]{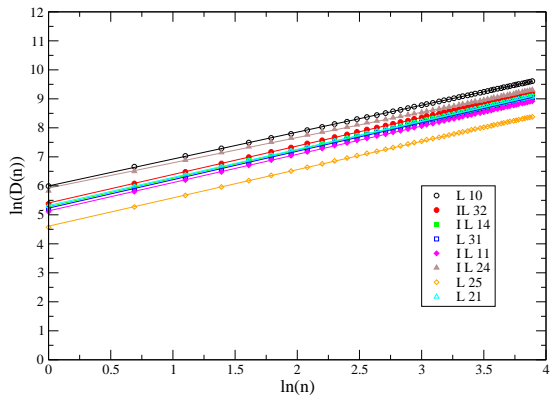}
\vskip 10mm
\includegraphics[width=.4\textwidth]{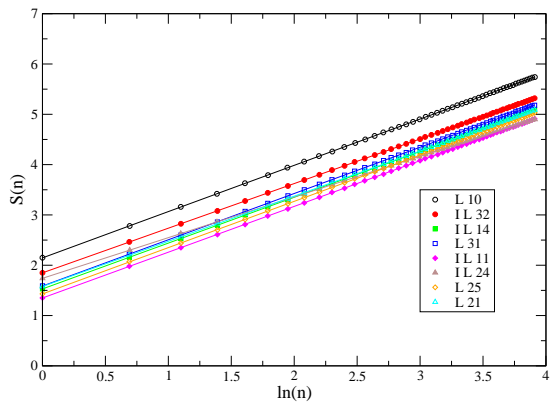}
\caption{Upper: Plot of ln $D(n)$ vs.  ln $n$ for subtrajectories $Z(n)$ corresponding to 8 representative subjects during saccadic eye movement. The legend indicates the correspondence between the symbol and the type/serial number of the subject. The exponents $H$ are found from the slope (see text) and tabulated in Table ~\ref{table2}. Lower:  Plot of Shannon entropies $S(n)$ vs. ln $n$ for saccades corresponding to 8 representative subjects. The legend indicates the correspondence between the symbol and the type/serial number of the subject. The exponents $\delta$ are found from the slope (see text) and tabulated in Table ~\ref{table2}.} 
\label{fig4}
\end{figure}

\begin{figure}
\centering
\includegraphics[width=.4\textwidth]{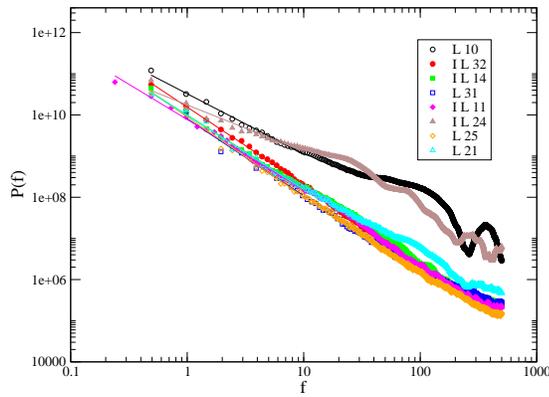}
\caption{Plot of the power spectrum $P(f)$ for eight subjects against frequency $f$. The legend indicates the correspondence between the symbol and the type/serial number of the subject. The exponents $\alpha$ are found from the slope (see text) and tabulated in Table ~\ref{table2}.}
\label{fig5}
\end{figure}

In the following, we analyze the spatial positions of the gazes of 8 representative subjects, spanning literates and illiterates, with an aim to determining the nature of saccadic dynamics; our results are presented in Figs. \ref{fig4} and \ref{fig5}, and summarised in Table \ref{table2}. For each of these subjects, a time series $x(t)$ is constructed from the Euclidean distances corresponding to the gaze coordinates at times   $t=1,2,...N$ during a given saccade.  Subtrajectories $Z_{i}(n)$ starting in the $i^{th}$ time bin and spanning a time duration $n$ are then computed using the relation $Z_{i}(t) = \sum_{j= 1 }^{n}[x(i+j-1)]$ \cite{stephen,scafetta1,scafetta2,scafetta3}; as the name implies, the subtrajectories $Z_{i}(n)$ are thus the cumulative Euclidean distances spanned in time windows of length $[i, i+n-1]$ during saccades. These subtrajectories form the basis of estimating the exponents $\delta$ and $H$ in what follows.

The standard deviation analysis method of \cite{scafetta1,scafetta3} provides the basis of the calculation of the Hurst exponent $H$. Defining the standard deviation $D(n)$ for subtrajectories of length $n$ :

\begin{equation}
D(n) = \sqrt{\frac{\sum_{i = 1 }^{N-n}[Z_{i}(n) - <Z(n)>]^{2}}{N-n-1}},
\label{eq.1}
\end{equation}

the Hurst exponent $H$ is obtained from $D(n) \propto n^{H}$ \cite{scafetta1,scafetta2,scafetta3}. The exponent $\delta$ is estimated using the diffusion entropy analysis (DEA) method of \cite{scafetta1,scafetta2}; this involves constructing the histograms of the end points of the subtrajectories $Z_{i}(n)$. If $p_{i}(n)$ is the probability that an end point of a subtrajectory $Z(n)$ of duration $n$ lies in the $i^{th}$ time bin, the Shannon entropy for a time duration $n$ is expressible as:

\begin{equation}
S(n)= -\sum_{i = 1 }^{N-m}p_{i} (n) ln (p_{i} (n)).
\label{eq.2}
\end{equation}

The DEA is based on the assumption of stationarity for the probability distribution function (pdf) characterising any time series as a diffusion process \cite{scafetta2, scafetta3}. If stationarity holds,  the pdf for our time series $x(t)$ can be written as $ p(x,t)=\frac{1}{t^{\delta}}F(\frac{x}{t^{\delta}})$. On going to the continuum limit of the Shannon entropy $S(t) =\displaystyle\int^{\infty}_{-\infty} p(x,t) ln(p(x,t))\,dx$, one gets $S(t) = A+ \delta ln(t)$, which yields for discrete time durations $n$:

\begin{equation}
S(n) = A+ \delta ln(n),
\label{eq.3}
\end{equation} 

where $A$ is some constant. Eqs. ~\ref{eq.2} and ~\ref{eq.3} were used to plot Fig. ~\ref{fig4}; the slopes of the lines in the upper and lower plots respectively yielded the values of the Hurst exponent $H$ and $\delta$ for eight representative subjects, which were then tabulated in Table \ref{table2}.
 
\begin{table}
\caption{Various exponents plotted for the sample subjects whose data are plotted in Figs. ~\ref{fig4} and ~\ref{fig5}.}
\label{table2}
\begin{center}
\begin{tabular}{l |l | l|l |l|l |}
\hline
Literacy Level & Serial No.&H & $\delta$ &$\alpha $ & $\mu$ \\ \hline
{Literates}&10&0.92 & 0.87& 1.46&1.85 \\ \cline{2-6}
&21 &0.96& 0.90&1.79&2.75 \\ \cline{2-6}
&25 &0.97&0.92&1.94&1.96 \\ \cline{2-6}
&31&0.98& 0.91&1.87&2.35 \\ \hline
{Illiterates}&11 &0.97& 0.90&1.71&2.81 \\ \cline{2-6}
& 14&0.98 & 0.89&1.78 &2.17\\ \cline{2-6}
&24 & 0.87&0.81&1.13 &2.14 \\ \cline{2-6}
&32 & 0.98&0.92&1.87&1.35\\ \hline
\end{tabular}
\end{center}
\end{table}

The values of $H$ ($H > 0.5$ indicates hyperdiffusion) obtained and the fact that $\delta \neq H $ across Table ~\ref{table2} both suggest that our data are consistent with Levy processes. Table ~\ref{table2} also contains values of the exponent $\mu$ obtained from  $P(x) = x^{-\mu}$ for each subject; the $\mu$ values so obtained independently confirm Levy-like behavior \cite{scafetta1,scafetta2}. Finally, another confirmation of hyperdiffusion comes from our measurement of the power spectrum $P(f) \sim 1/f^{\alpha}$, plotted against frequency $f$ for all 8 subjects (Fig.~\ref{fig5}). The resulting values of the exponent $\alpha$ (see Table ~\ref{table2}) all lie between 1 and 2, and are consistent with the nontrivial temporal correlations typical of hyperdiffusive dynamics \cite{Aks}.

Having shown the internal consistency of our results, we now report external support. The findings of Van Loon et al. \cite{VanLoon} who found non-Gaussian behavior for the timing of sequence for the '\ later '\  saccades in a multiple-fixation search like our own, buttress our conclusions. They are in contrast to the findings of \cite{Mark}, where fractional Brownian behavior was observed in the predictive eye movements of saccades; this is probably because, in our case, the experimental paradigm is \textit{not} predictive at all, and participants are \textit{not} aware that they are performing a task. Finally, our results support the findings of \cite{Reynolds}, where it was shown that fractional Levy-like dynamics (rather than ordinary Levy walks with $H = 0.5$) predominate in foraging, a situation in which animals are typically unaware of where next they will find food. This is very similar to what obtains in our case: participants are unaware of the locations of their next targets, and accordingly, the Hurst exponent $H$ takes on values that are much greater than $0.5$ (see Table ~\ref{table2}). Our tentative suggestion is therefore that  saccadic eye movements obey fractional Levy dynamics, irrespective of the literacy level of the subject.

In conclusion, our most striking observation here is that there is a clear separation of behavior in visual processing, via the distributions of saccadic and fixation times; the former shows clear scale-invariance, while the latter does not. The effect of literacy shows up only in the fixation dynamics, where literate subjects typically show the presence of an attentional timescale around $220$ msecs on average. However, the saccades of \textit{both} literate and illiterate subjects appear to  be characterised by fractional Levy-like dynamics; we suggest that this might be explicable by analogies with foraging dynamics in animals \cite{Rhodes}, since in both cases the subjects are completely unaware of the locations of their next targets.

\acknowledgments
We thank Niharika Singh for helping with stimuli preparation and data collection. DPS was partially supported by a grant awarded to AM from the Department of Science and Technology, India, as part of the Cognitive Science Initiative. RKM thanks the Centre of Behavioural and Cognitive Science, Allahabad for an internal grant.


\begin{thebibliography}{0}
\bibitem{littest} 
We administered a reading test to confirm the literacy level of our
participants. Out of 96 Hindi words, literates on average read aloud 93.8 words correctly (with standard deviation  2.04) whereas  illiterates on average
only read aloud 6.5 words correctly (with standard deviation 7.77).
\end{thebibliography}
\end{document}